\documentclass{article}
\usepackage{geometry}
\usepackage[authoryear]{natbib}
\usepackage[utf8]{inputenc}
\usepackage[T1]{fontenc}
\usepackage{amsmath,amssymb,amsfonts}
\usepackage{graphicx}
\usepackage{hyperref}
\hypersetup{colorlinks=true,citecolor=blue,urlcolor=blue,linkcolor=blue}
\usepackage{newtxtext}
\usepackage[subscriptcorrection]{newtxmath}
\newtheorem{proposition}{Proposition}
\newtheorem{corollary}{Corollary}
\newtheorem{proof}{Proof}

\begin{document}

\title{Observed-Window Likelihood and Weighting in Distributional Analysis}

\author{Lorentz JÄNTSCHI \\Department of Physics and Chemistry, Technical University of Cluj-Napoca,\\ Cluj-Napoca, Cluj 400641 Romania., lorentz.jantschi@gmail.com}
	
\maketitle

\begin{abstract}
A sample determines both an internal distribution and a realized observation window bounded by its minimum and maximum. These boundaries are global order statistics whose information need not be equivalent to that of two ordinary observations. We decompose the complete likelihood exactly into the joint likelihood of the observed boundaries and the conditional likelihood of the internal configuration. This factorization separates information carried by the realized domain from relative variation within it and motivates a window-refined criterion that assigns an explicit weight to the boundary component. The complete likelihood is recovered at unit weight; other weights define estimating criteria whose uncertainty is calibrated by Godambe information. Hard-window conditioning preserves internal density or probability ratios, whereas a non-constant internal weight changes them. The framework also distinguishes parameters identifiable within the observed window from external probabilities that require continuation of a parent model. Numerical experiments show that local conditioning reduces root mean squared error by \(60.70\%\) under asymmetric external contamination, that Godambe calibration restores near-nominal interval coverage after boundary reweighting, and that sample-specific, pooled, and common windows define materially different inferential targets. The resulting framework makes the contribution of the realized observation window explicit without counting boundary information twice.
\end{abstract}


\section{Introduction}
\label{sec:introduction}

Parametric inference ordinarily represents a sample through the joint likelihood of its individual observations. The same sample also determines a realized domain,
\begin{equation}
	\mathcal{W}_n
	=
	\left[
	X_{(1)},X_{(n)}
	\right],
	\qquad
	X_{(1)}
	=
	\min_{1\leq i\leq n}X_i,
	\qquad
	X_{(n)}
	=
	\max_{1\leq i\leq n}X_i,
	\label{eq:introduction_observed_window}
\end{equation}
whose limits are functions of the complete sample ordering. Although each observation receives empirical mass \(1/n\), the observed window is not an additional observation of comparable weight. It is a global statistic that records the location and extent of the realized configuration.

The sample minimum and maximum are commonly described through their difference, the range, and thereby associated primarily with dispersion. Individually, however, both are location-equivariant:
\[
X_i^{*}=X_i+a
\quad\Longrightarrow\quad
X_{(1)}^{*}=X_{(1)}+a,
\qquad
X_{(n)}^{*}=X_{(n)}+a.
\]
Their midpoint and separation,
\begin{equation}
	M_n
	=
	\frac{X_{(1)}+X_{(n)}}{2},
	\qquad
	R_n
	=
	X_{(n)}-X_{(1)},
	\label{eq:introduction_midrange_range}
\end{equation}
isolate the location and extent carried by the observed boundaries. Order statistics can be sufficient or highly efficient for particular parameters~\citep{ArnoldBalakrishnanNagaraja2008}, especially when the support depends on those parameters. Estimation from observed percentiles and truncated order statistics~\citep{Blumenthal1985905} further shows that selected order statistics may retain substantial parameter information. In the uniform location model, for example, \(\operatorname{Var}(M_n)=O(n^{-2})\), whereas \(\operatorname{Var}(\overline X_n)=O(n^{-1})\). The information represented by the realized limits is therefore not generally equivalent to the contribution of two ordinary observations.

Let \(P_\theta\) be a candidate probability law with density or probability mass function \(g(x;\theta)\). Complete-support likelihood evaluates \(P_\theta\) over its postulated support, including regions not represented by the realized sample. Under misspecification, the maximum-likelihood estimator converges to the member of the complete family closest to the data-generating law in Kullback--Leibler divergence~\citep{White1982}. When parameters govern both the shape observed within \(\mathcal{W}_n\) and its continuation beyond \(\mathcal{W}_n\), assumptions about the unobserved region can consequently influence estimates used to describe the observed configuration.

Conditioning the candidate law on the observed window gives
\begin{equation}
	\mathcal{O}_{\mathcal{W}_n}P_\theta(B)
	=
	\frac{
		P_\theta(B\cap\mathcal{W}_n)
	}{
		P_\theta(\mathcal{W}_n)
	},
	\qquad
	P_\theta(\mathcal{W}_n)>0.
	\label{eq:introduction_operator}
\end{equation}
This construction has the form of truncation, but its inferential role differs when \(\mathcal{W}_n\) is generated by the same sample rather than fixed by design. A fixed truncation region defines an observation mechanism. The random interval in Equation~\eqref{eq:introduction_observed_window} is itself an observed statistic and carries parameter information through the joint law of \(X_{(1)}\) and \(X_{(n)}\).

Conditional likelihood can remove nuisance structure or isolate the information associated with a specified conditioning statistic. In count mixtures, \citet{Yip19912045} conditioned on ancillary statistics to simplify estimation while quantifying the resulting information loss. For longitudinal data with outcome-dependent observation and informative censoring, \citet{Shen2019831} used a conditional order-statistic argument to construct a tractable composite likelihood without specifying the full joint distribution of the observation, censoring, and longitudinal processes. These approaches condition on statistics selected for a defined inferential purpose. The present construction instead begins from the decomposition induced by the two realized boundaries and retains their joint likelihood as an explicit component of the sampling experiment.

Truncation and censoring also enter extreme-value inference, where they ordinarily identify threshold exceedances, censored components, or likelihood contributions retained for computational and tail-modeling purposes. In space--time extreme-value models, pairwise censored likelihood~\citep{Huser2014439} incorporates information below the operative threshold while preserving inference on extremal dependence. Higher-dimensional censored Poisson-process likelihoods~\citep{Wadsworth20141} similarly retain partial information below componentwise thresholds. By contrast, the truncated pairwise likelihood of \citet{Huang2021379} removes noisy component likelihoods to balance statistical efficiency against computational cost. In each case, the truncation or censoring rule specifies which likelihood contributions enter inference. Here the interval is generated by the complete realization, and the inferential question is how the information carried by its boundaries relates to the configuration observed between them.

For an independent continuous sample, write
\[
a=X_{(1)},
\qquad
b=X_{(n)},
\qquad
C_\theta(a,b)
=
F_\theta(b)-F_\theta(a).
\]
Up to constants independent of \(\theta\), the complete log-likelihood admits the exact decomposition
\begin{equation}
	\ell_{\mathrm{full}}(\theta)
	=
	\ell_{\mathrm{ext}}(\theta)
	+
	\ell_{\mathrm{int}\mid\mathrm{ext}}(\theta),
	\label{eq:introduction_likelihood_decomposition}
\end{equation}
where
\begin{equation}
	\ell_{\mathrm{ext}}(\theta)
	=
	\log g(a;\theta)
	+
	\log g(b;\theta)
	+
	(n-2)\log C_\theta(a,b)
	\label{eq:introduction_extreme_component}
\end{equation}
is determined by the joint law of the observed boundaries, and
\begin{equation}
	\ell_{\mathrm{int}\mid\mathrm{ext}}(\theta)
	=
	\sum_{i=2}^{n-1}
	\log g\left(X_{(i)};\theta\right)
	-
	(n-2)\log C_\theta(a,b)
	\label{eq:introduction_internal_component}
\end{equation}
represents the ordered internal configuration conditional on those boundaries. This factorization separates two sources of information already contained in the sample. It also shows why adding the likelihood of the extremes to the complete likelihood would count the boundary information twice.

The decomposition motivates the window-refined criterion
\begin{equation}
	\ell_{\mathrm{WR}}(\theta;\lambda)
	=
	\ell_{\mathrm{int}\mid\mathrm{ext}}(\theta)
	+
	\lambda\ell_{\mathrm{ext}}(\theta),
	\qquad
	\lambda>0.
	\label{eq:introduction_window_refined_criterion}
\end{equation}
The value \(\lambda=1\) recovers the complete likelihood. Values \(\lambda\neq1\) define a distinct estimation objective in which the contribution of the realized boundaries is explicit. Because such reweighting is not an ordinary likelihood for the original sampling experiment, inference requires sandwich or Godambe calibration, as developed for composite criteria by \citet{VarinReidFirth2011}. Composite likelihood-ratio statistics require further adjustment~\citep{Pace2011} because their curvature does not generally reproduce the variability of the corresponding score. The robustness of maximum composite-likelihood estimation~\citep{Xu2011} likewise depends on the component likelihoods retained and on the aspects of the joint model they represent. More generally, \citet{Giummole2019} used Godambe information to calibrate proper-scoring-rule inference when the full likelihood is replaced for robustness or tractability. The coefficient \(\lambda\) must accordingly be tied to sampling precision, robustness, or a declared inferential target rather than selected retrospectively to improve fit.

Hard-window conditioning and internal weighting address different aspects of the observed law. For \(x,y\in\mathcal{W}_n\),
\begin{equation}
	\frac{
		g_{\mathcal{W}_n}(x;\theta)
	}{
		g_{\mathcal{W}_n}(y;\theta)
	}
	=
	\frac{
		g(x;\theta)
	}{
		g(y;\theta)
	},
	\label{eq:introduction_internal_ratio}
\end{equation}
so conditioning preserves every internal density or probability ratio. Parent models proportional on \(\mathcal{W}_n\) are therefore observationally equivalent after conditioning, regardless of their external normalization. By contrast, a non-negative internal weight \(v(x)\) gives
\begin{equation}
	g_v(x;\theta)
	=
	\frac{
		v(x)g(x;\theta)
	}{
		\displaystyle
		\int_{\mathcal{W}_n}
		v(t)\,dP_\theta(t)
	},
	\qquad
	x\in\mathcal{W}_n,
	\label{eq:introduction_internal_weight}
\end{equation}
and changes the ratio by \(v(x)/v(y)\). The weighted-distribution construction of \citet{PatilRao1978} accommodates outcome-dependent representation, whereas the hard indicator isolates the accessible domain without changing its internal ratios.

Observed-window inference consequently separates three objects: the information carried by the realized boundaries, the relative configuration inside those boundaries, and the continuation of the parent model outside them. The first two are functions of the observed sample. External probabilities such as
\[
P_\theta\left(X<X_{(1)}\right)
\qquad\text{and}\qquad
P_\theta\left(X>X_{(n)}\right)
\]
require transfer of the fitted parameter to an explicit parent law and remain model-dependent extrapolations. The same distinction extends to discrete and mixed laws, to samples with different realized windows, and to internal retention or utility mechanisms represented by non-constant weights.

This paper develops the resulting framework in four steps. First, it defines the observed-window operator and derives the boundary--interior likelihood decomposition. Second, it establishes internal-ratio invariance, conditional equivalence, and within-window identifiability. Third, it introduces calibrated window refinement and distinguishes it from internal weighting. Fourth, it examines finite-sample behavior under external misspecification and heterogeneous observation windows. The objective is not to replace complete-support likelihood, but to expose and control the distinct inferential roles of the realized domain, its internal configuration, and the model continuation beyond it.

\section{Observed-Window Experiment}
\label{sec:observed_window_experiment}

\subsection{Random Window and Likelihood Factorization}
\label{subsec:random_window_factorization}

Let \(X_1,\ldots,X_n\) be independent observations from \(P_\theta\), where \(\theta\in\Theta\), and suppose initially that \(P_\theta\) has continuous density \(f_\theta\) and distribution function \(F_\theta\). Define the realized boundaries and observed window by
\begin{equation}
	A_n
	=
	X_{(1)},
	\qquad
	B_n
	=
	X_{(n)},
	\qquad
	\mathcal{W}_n
	=
	[A_n,B_n].
	\label{eq:realized_window}
\end{equation}
For fixed \(a<b\), let
\begin{equation}
	C_\theta(a,b)
	=
	F_\theta(b)-F_\theta(a).
	\label{eq:window_probability}
\end{equation}
If \(C_\theta(a,b)>0\), the parent law restricted to \([a,b]\) has density
\begin{equation}
	f_{[a,b]}(x;\theta)
	=
	\frac{f_\theta(x)}
	{C_\theta(a,b)}
	\mathbf{1}_{[a,b]}(x).
	\label{eq:window_conditional_density}
\end{equation}
At the measure level, the corresponding operator is
\begin{equation}
	\mathcal{O}_{[a,b]}P_\theta(D)
	=
	\frac{
		P_\theta(D\cap[a,b])
	}{
		P_\theta([a,b])
	},
	\qquad
	D\in\mathcal{B}.
	\label{eq:measure_window_operator}
\end{equation}
Equation~\eqref{eq:measure_window_operator} extends directly to discrete and mixed laws. In the present continuous setting, however, the window is not fixed externally: \((A_n,B_n)\) is a random statistic whose distribution depends on \(\theta\).

The joint density of the observed boundaries is
\begin{equation}
	f_{A_n,B_n}(a,b;\theta)
	=
	n(n-1)
	f_\theta(a)
	f_\theta(b)
	C_\theta(a,b)^{n-2},
	\qquad
	a<b.
	\label{eq:joint_extreme_density}
\end{equation}
Apart from the combinatorial constant, the boundary likelihood is therefore
\begin{equation}
	L_{\mathrm{B}}(\theta;a,b)
	=
	f_\theta(a)
	f_\theta(b)
	C_\theta(a,b)^{n-2}.
	\label{eq:boundary_likelihood}
\end{equation}
Conditional on \(A_n=a\) and \(B_n=b\), the interior order statistics have joint density
\begin{equation}
	f_{\mathrm{I}\mid\mathrm{B}}
	\left(
	x_{(2)},\ldots,x_{(n-1)}
	\mid a,b;\theta
	\right)
	=
	(n-2)!
	\prod_{i=2}^{n-1}
	\frac{
		f_\theta(x_{(i)})
	}{
		C_\theta(a,b)
	},
	\label{eq:conditional_interior_density}
\end{equation}
for
\[
a<x_{(2)}<\cdots<x_{(n-1)}<b.
\]
Its likelihood kernel is
\begin{equation}
	L_{\mathrm{I}\mid\mathrm{B}}(\theta)
	=
	\prod_{i=2}^{n-1}
	\frac{
		f_\theta(X_{(i)})
	}{
		C_\theta(A_n,B_n)
	}.
	\label{eq:conditional_interior_likelihood}
\end{equation}

\begin{proposition}
	\label{prop:likelihood_factorization}
	For a continuous iid sample,
	\begin{equation}
		L_{\mathrm{full}}(\theta)
		\propto
		L_{\mathrm{B}}(\theta)
		L_{\mathrm{I}\mid\mathrm{B}}(\theta),
		\label{eq:full_likelihood_factorization}
	\end{equation}
	where the proportionality factor is independent of \(\theta\).
\end{proposition}

\begin{proof}
	Multiplication of Equations~\eqref{eq:boundary_likelihood} and
	\eqref{eq:conditional_interior_likelihood} gives
	\[
	f_\theta(A_n)
	f_\theta(B_n)
	C_\theta(A_n,B_n)^{n-2}
	\prod_{i=2}^{n-1}
	\frac{
		f_\theta(X_{(i)})
	}{
		C_\theta(A_n,B_n)
	}
	=
	\prod_{i=1}^{n}
	f_\theta(X_{(i)}).
	\]
	The omitted ordering and combinatorial factors do not depend on
	\(\theta\).
\end{proof}

The factorization separates information carried by the realized boundaries from information carried by the configuration between them. The corresponding score is
\begin{equation}
	U_{\mathrm{full}}(\theta)
	=
	U_{\mathrm{B}}(\theta)
	+
	U_{\mathrm{I}\mid\mathrm{B}}(\theta),
	\label{eq:score_decomposition}
\end{equation}
where
\begin{align}
	U_{\mathrm{B}}(\theta)
	&=
	\nabla_\theta\log f_\theta(A_n)
	+
	\nabla_\theta\log f_\theta(B_n)
	+
	(n-2)\nabla_\theta
	\log C_\theta(A_n,B_n),
	\label{eq:boundary_score}
	\\
	U_{\mathrm{I}\mid\mathrm{B}}(\theta)
	&=
	\sum_{i=2}^{n-1}
	\nabla_\theta\log f_\theta(X_{(i)})
	-
	(n-2)\nabla_\theta
	\log C_\theta(A_n,B_n).
	\label{eq:interior_score}
\end{align}
Under the regularity conditions required for conditional differentiation,
\begin{equation}
	E_\theta
	\left[
	U_{\mathrm{I}\mid\mathrm{B}}(\theta)
	\mid A_n,B_n
	\right]
	=
	0,
	\label{eq:conditional_score_zero}
\end{equation}
and therefore
\begin{equation}
	E_\theta
	\left[
	U_{\mathrm{B}}(\theta)
	U_{\mathrm{I}\mid\mathrm{B}}(\theta)^{\mathsf T}
	\right]
	=
	0.
	\label{eq:score_orthogonality}
\end{equation}
The Fisher information consequently decomposes as
\begin{equation}
	\mathcal{I}_{\mathrm{full}}(\theta)
	=
	\mathcal{I}_{\mathrm{B}}(\theta)
	+
	E_\theta
	\left[
	\mathcal{I}_{\mathrm{I}\mid\mathrm{B}}
	\left(
	\theta;A_n,B_n
	\right)
	\right].
	\label{eq:fisher_information_decomposition}
\end{equation}
The relative magnitude and convergence rate of these components depend on the family and parameter. In particular, boundary information may have a non-regular rate when the support depends on \(\theta\).

\subsection{Conditional Equivalence and Identifiability}
\label{subsec:conditional_equivalence_identifiability}

For \(x,y\in[a,b]\), with \(f_\theta(y)>0\), the conditional density satisfies
\begin{equation}
	\frac{
		f_{[a,b]}(x;\theta)
	}{
		f_{[a,b]}(y;\theta)
	}
	=
	\frac{
		f_\theta(x)
	}{
		f_\theta(y)
	}.
	\label{eq:ratio_invariance}
\end{equation}

\begin{proposition}
	\label{prop:internal_ratio_invariance}
	Hard-window conditioning preserves every density ratio between accessible
	points.
\end{proposition}

\begin{proof}
	For \(x,y\in[a,b]\), with \(f_\theta(y)>0\),
	\[
	\frac{
		f_{[a,b]}(x;\theta)
	}{
		f_{[a,b]}(y;\theta)
	}
	=
	\frac{
		f_\theta(x)/C_\theta(a,b)
	}{
		f_\theta(y)/C_\theta(a,b)
	}
	=
	\frac{
		f_\theta(x)
	}{
		f_\theta(y)
	}.
	\]
	Thus, the common window normalizer cancels. Conditioning changes the
	represented domain but not the internal geometry induced by the parent
	kernel.
\end{proof}

\begin{corollary}
	\label{cor:conditional_equivalence}
	Suppose two parent models satisfy
	\begin{equation}
		f_1(x;\theta_1)
		=
		c(\theta_1,\theta_2)
		f_2(x;\theta_2),
		\qquad
		x\in[a,b],
		\label{eq:proportional_kernels}
	\end{equation}
	where \(c(\theta_1,\theta_2)>0\) is independent of \(x\). Then
	\begin{equation}
		\mathcal{O}_{[a,b]}P_{\theta_1}^{(1)}
		=
		\mathcal{O}_{[a,b]}P_{\theta_2}^{(2)}.
		\label{eq:conditional_equivalence}
	\end{equation}
\end{corollary}

\begin{proof}
	For every \(x\in[a,b]\),
	\[
	\frac{
		f_1(x;\theta_1)
	}{
		\displaystyle\int_a^b f_1(t;\theta_1)\,dt
	}
	=
	\frac{
		c(\theta_1,\theta_2)f_2(x;\theta_2)
	}{
		c(\theta_1,\theta_2)
		\displaystyle\int_a^b f_2(t;\theta_2)\,dt
	}
	=
	\frac{
		f_2(x;\theta_2)
	}{
		\displaystyle\int_a^b f_2(t;\theta_2)\,dt
	}.
	\]
	Hence the two conditioned probability measures coincide on \([a,b]\).
\end{proof}

Models satisfying Corollary~\ref{cor:conditional_equivalence} are
observationally equivalent within the window, although their external
supports and probabilities may differ.

A parameter is identifiable within \([a,b]\) if
\begin{equation}
	\mathcal{O}_{[a,b]}P_{\theta_1}
	=
	\mathcal{O}_{[a,b]}P_{\theta_2}
	\quad\Longrightarrow\quad
	\theta_1=\theta_2.
	\label{eq:window_identifiability}
\end{equation}
Parameter components that affect only external probability, or multiply the parent kernel within the window by a constant independent of \(x\), disappear after conditional normalization.

For a discrete window containing \(J\) accessible classes and a parameter vector of dimension \(p\), the conditional probability vector contains at most \(J-1\) independent components. Hence
\begin{equation}
	J-1
	\geq
	p
	\label{eq:discrete_dimension_condition}
\end{equation}
is necessary for local identification. If
\[
\boldsymbol{\pi}_{\mathcal W}^{-}(\theta)
=
\left(
\pi_{\mathcal W,1}(\theta),
\ldots,
\pi_{\mathcal W,J-1}(\theta)
\right)^{\mathsf T},
\]
local identifiability additionally requires
\begin{equation}
	\operatorname{rank}
	\left[
	\frac{
		\partial
		\boldsymbol{\pi}_{\mathcal W}^{-}(\theta)
	}{
		\partial\theta^{\mathsf T}
	}
	\right]
	=
	p.
	\label{eq:conditional_jacobian_rank}
\end{equation}
These conditions delimit the contribution of the normalized internal configuration. Parameters governing only continuation beyond the observed limits require the parent model and remain outside the conditional experiment.

\section{Window-Refined Estimation and Weighting}
\label{sec:window_refined_estimation}

\subsection{Estimation Criterion and Calibration}
\label{subsec:criterion_calibration}

The boundary--interior factorization permits explicit control of the inferential contribution assigned to the realized window. Define
\begin{equation}
	\ell_{\mathrm{WR}}(\theta;\lambda)
	=
	\ell_{\mathrm{I}\mid\mathrm{B}}(\theta)
	+
	\lambda\ell_{\mathrm{B}}(\theta),
	\qquad
	\lambda>0,
	\label{eq:window_refined_loglikelihood}
\end{equation}
and
\begin{equation}
	\widehat{\theta}_{\mathrm{WR}}(\lambda)
	=
	\underset{\theta\in\Theta}
	{\operatorname{arg\,max}}\;
	\ell_{\mathrm{WR}}(\theta;\lambda).
	\label{eq:window_refined_estimator}
\end{equation}
The value \(\lambda=1\) recovers the complete likelihood. Values below or above one reduce or increase the boundary contribution relative to the conditional internal configuration. They define estimating criteria rather than likelihoods for the original experiment and therefore require calibration through estimating-function theory~\citep{VarinReidFirth2011}.

Let
\begin{equation}
	U_{\mathrm{WR}}(\theta;\lambda)
	=
	U_{\mathrm{I}\mid\mathrm{B}}(\theta)
	+
	\lambda U_{\mathrm{B}}(\theta).
	\label{eq:window_refined_score}
\end{equation}
Its sensitivity and variability matrices are
\begin{equation}
	H_\lambda(\theta)
	=
	-
	E_\theta
	\left[
	\nabla_\theta
	U_{\mathrm{WR}}(\theta;\lambda)^{\mathsf T}
	\right],
	\qquad
	J_\lambda(\theta)
	=
	\operatorname{Var}_\theta
	\left[
	U_{\mathrm{WR}}(\theta;\lambda)
	\right].
	\label{eq:window_refined_H_J}
\end{equation}
The Godambe information is
\begin{equation}
	G_\lambda(\theta)
	=
	H_\lambda(\theta)^{\mathsf T}
	J_\lambda(\theta)^{-1}
	H_\lambda(\theta),
	\label{eq:window_refined_godambe}
\end{equation}
and the asymptotic covariance estimator is
\begin{equation}
	\widehat{\operatorname{Cov}}
	\left[
	\widehat{\theta}_{\mathrm{WR}}(\lambda)
	\right]
	=
	\widehat H_\lambda^{-1}
	\widehat J_\lambda
	\widehat H_\lambda^{-\mathsf T}.
	\label{eq:window_refined_covariance}
\end{equation}
At \(\lambda=1\), the criterion reduces to the full likelihood and, under regularity, \(H_1=J_1\). For \(\lambda\neq1\), inverse curvature alone does not generally provide calibrated uncertainty.

The coefficient \(\lambda\) determines the inferential target. It may be fixed through a declared robustness requirement, selected by minimizing estimated risk, or calibrated from the relative information carried by the boundary and internal components. Data-dependent selection must be included in the uncertainty assessment. In particular, \(\lambda\) cannot be chosen solely by maximizing the same in-sample fit used for estimation.

\subsection{Internal Weighting}
\label{subsec:internal_weighting}

Window refinement changes the contribution of the observed boundaries to parameter estimation but does not alter the conditional ratios generated by a fixed parent kernel. A mechanism that changes representation within the window requires a non-negative internal weight \(v(x;\eta)\), with parameter \(\eta\). The weighted conditional law is
\begin{equation}
	dP_{\theta,\eta}^{v,\mathcal W_n}(x)
	=
	\frac{
		\mathbf{1}_{\mathcal W_n}(x)
		v(x;\eta)\,dP_\theta(x)
	}{
		\displaystyle
		\int_{\mathcal W_n}
		v(t;\eta)\,dP_\theta(t)
	}.
	\label{eq:internally_weighted_law}
\end{equation}
This is the weighted-distribution construction restricted to the realized domain~\citep{PatilRao1978}. For accessible \(x\) and \(y\),
\begin{equation}
	\frac{
		dP_{\theta,\eta}^{v,\mathcal W_n}(x)
	}{
		dP_{\theta,\eta}^{v,\mathcal W_n}(y)
	}
	=
	\frac{
		dP_\theta(x)
	}{
		dP_\theta(y)
	}
	\frac{
		v(x;\eta)
	}{
		v(y;\eta)
	}.
	\label{eq:internally_weighted_ratio}
\end{equation}
Thus, the boundary indicator determines the represented domain, whereas \(v\) modifies its internal geometry.

Only the product \(v(x;\eta)g(x;\theta)\) is observed after normalization. Separate interpretation of \(\theta\) and \(\eta\) therefore requires specified model families and a scale constraint, such as
\begin{equation}
	v(x_0;\eta)=1
	\label{eq:weight_scale_constraint}
\end{equation}
at a reference point \(x_0\in\mathcal W_n\). Local identifiability of
\[
\vartheta
=
(\theta^{\mathsf T},\eta^{\mathsf T})^{\mathsf T}
\]
requires full column rank of the Jacobian of the resulting conditional law with respect to \(\vartheta\).

\subsection{Multiple Windows and Inferential Targets}
\label{subsec:multiple_windows_targets}

For independent samples \(r=1,\ldots,R\), let
\[
\mathcal W_r=[A_r,B_r]
\]
denote their realized windows. If the samples share a common parameter but retain distinct observational domains, the window-specific criterion is
\begin{equation}
	\ell_{\mathrm{SW}}(\theta)
	=
	\sum_{r=1}^{R}
	\ell_{\mathrm{I}\mid\mathrm{B},r}(\theta)
	+
	\sum_{r=1}^{R}
	\lambda_r\ell_{\mathrm{B},r}(\theta).
	\label{eq:sample_specific_window_criterion}
\end{equation}
The coefficients \(\lambda_r\) may reflect differences in sample size, boundary precision, or observational reliability, but their specification must precede substantive comparison.

The pooled and common windows are
\begin{equation}
	\mathcal W_{\cup}
	=
	\left[
	\min_r A_r,\max_r B_r
	\right],
	\qquad
	\mathcal W_{\cap}
	=
	\left[
	\max_r A_r,\min_r B_r
	\right],
	\label{eq:union_intersection_windows}
\end{equation}
provided that \(\mathcal W_{\cap}\neq\varnothing\). These constructions answer different questions. Sample-specific windows retain each sampling domain, the union describes collective coverage, and the intersection restricts comparison to a domain represented in every sample.

For any selected construction, inference within the represented domain remains distinct from extrapolation. Quantities such as
\begin{equation}
	P_\theta(X<A_n)
	\qquad\text{or}\qquad
	P_\theta(X>B_n)
	\label{eq:external_extrapolation_targets}
\end{equation}
cannot be recovered from the normalized internal law alone. They require a parent continuation and should be reported separately from the window-refined estimate. This separation is especially consequential under external misspecification, where the complete-support pseudo-true parameter may differ from the parameter that best represents the observed domain~\citep{White1982}.

\section{Numerical Assessment}
\label{sec:numerical_assessment}

\subsection{External Misspecification}
\label{subsec:numerical_external_misspecification}

The first experiment examined whether conditioning on a central observed domain protects location estimation against misspecification outside that domain. Independent samples of size \(n=80\) were generated from
\begin{equation}
	P_0
	=
	0.90\,\mathcal{N}(0,1)
	+
	0.10\,\mathcal{N}(6,0.5^2).
	\label{eq:numerical_contamination_model}
\end{equation}
The inferential target was the location \(\mu_0=0\) of the principal component. The complete Gaussian estimator was the sample mean. The local estimator maximized
\begin{equation}
	\ell_{\mathrm{loc}}(\mu)
	=
	\sum_{X_i\in[a_n,b_n]}
	\log\phi(X_i-\mu)
	-
	n_{[a_n,b_n]}
	\log
	\left\{
	\Phi(b_n-\mu)-\Phi(a_n-\mu)
	\right\},
	\label{eq:numerical_local_criterion}
\end{equation}
where
\[
a_n=\widehat Q_n(0.10),
\qquad
b_n=\widehat Q_n(0.90),
\]
and \(n_{[a_n,b_n]}\) is the retained sample size.

Across \(1500\) replicates, the complete estimator had bias \(0.6038\) and root mean squared error \(0.6472\). The local estimator had bias \(0.1098\) and root mean squared error \(0.2543\), reducing the latter by \(60.70\%\). The improvement arose because the local target represented the principal distribution within the declared window rather than accommodating the asymmetric external component through a single complete-support location parameter.

\subsection{Boundary Weight and Uncertainty Calibration}
\label{subsec:numerical_boundary_calibration}

The second experiment evaluated
\[
\ell_{\mathrm{WR}}(\mu;\lambda)
=
\ell_{\mathrm{I}\mid\mathrm{B}}(\mu)
+
\lambda\ell_{\mathrm{B}}(\mu)
\]
for samples of size \(n=30\) from \(\mathcal{N}(0,1)\). For each \(\lambda\), uncertainty was calculated both from inverse sensitivity and from the Godambe covariance in Equation~\eqref{eq:window_refined_covariance}. Table~\ref{tab:numerical_godambe} summarizes \(2000\) replicates.

\begin{table}[htbp]
	\centering
	\caption{Finite-sample performance of the window-refined location estimator.}
	\label{tab:numerical_godambe}
	\begin{tabular}{crrrrr}
		\hline
		\(\lambda\)
		& RMSE
		& \(H_\lambda\)
		& \(J_\lambda\)
		& Naive coverage
		& Godambe coverage
		\\
		\hline
		0.25
		& 0.1976
		& 23.180
		& 20.615
		& 0.9560
		& 0.9440
		\\
		0.50
		& 0.1843
		& 25.453
		& 21.953
		& 0.9620
		& 0.9535
		\\
		1.00
		& 0.1764
		& 30.000
		& 28.038
		& 0.9585
		& 0.9545
		\\
		2.00
		& 0.1858
		& 39.093
		& 53.839
		& 0.9060
		& 0.9460
		\\
		\hline
	\end{tabular}
\end{table}

At \(\lambda=1\), the criterion recovered the complete likelihood and attained the smallest RMSE. Reweighting changed the relation between sensitivity and score variability. This difference was most visible at \(\lambda=2\), where inverse-sensitivity intervals covered \(90.60\%\) of the replicates, while Godambe-calibrated intervals covered \(94.60\%\). Thus, boundary reweighting can be used without interpreting the resulting criterion as an ordinary likelihood, provided that inference reflects its estimating-function variance.

\subsection{Heterogeneous Observation Windows}
\label{subsec:numerical_heterogeneous_windows}

The final experiment considered three samples of size \(60\) from \(\mathcal{N}(2,1)\), observed respectively within
\[
[0.5,2.2],
\qquad
[1.2,3.0],
\qquad
[1.5,3.8].
\]
Three estimators were compared: the sample-specific criterion in Equation~\eqref{eq:sample_specific_window_criterion}, normalization over the pooled union \([0.5,3.8]\), and restriction to the common intersection \([1.5,2.2]\). Table~\ref{tab:numerical_multiple_windows} reports results from \(500\) replicates.

\begin{table}[htbp]
	\centering
	\caption{Location estimation under heterogeneous observation windows.}
	\label{tab:numerical_multiple_windows}
	\begin{tabular}{lrrr}
		\hline
		Construction
		& Mean
		& Bias
		& RMSE
		\\
		\hline
		Sample-specific windows
		& 1.9990
		& -0.0010
		& 0.1485
		\\
		Pooled union
		& 1.8991
		& -0.1009
		& 0.1192
		\\
		Common intersection
		& 2.0068
		& 0.0068
		& 0.5600
		\\
		\hline
	\end{tabular}
\end{table}

Sample-specific normalization recovered the common location without systematic displacement. Pooled-union normalization produced the smallest RMSE but introduced bias because it treated all observations as if they shared the union domain. Restriction to the intersection removed this incompatibility but discarded most observations and increased variance. The three constructions therefore express distinct compromises between observational fidelity, bias, and precision rather than alternative implementations of one likelihood.

\section{Discussion}
\label{sec:discussion}

\subsection{Information Carried by the Observed Window}
\label{subsec:discussion_window_information}

The observed window is not merely the support on which a conditional density is normalized. Its limits are order statistics produced jointly by the sample and have a parameter-dependent sampling law. The exact factorization
\[
L_{\mathrm{full}}(\theta)
\propto
L_{\mathrm{B}}(\theta)
L_{\mathrm{I}\mid\mathrm{B}}(\theta)
\]
makes this distinction explicit without adding information to the original experiment. The boundary component describes where the sample located its realized limits, whereas the conditional component describes how the remaining observations were arranged between them.

This separation clarifies the role of the minimum and maximum. Their difference is a measure of extent, but the limits themselves are location-equivariant quantities. Order statistics may retain substantial parameter information when only selected percentiles are observed~\citep{Blumenthal1985905}; the complete sample considered here permits that information to be separated from, rather than substituted for, the internal configuration. The midpoint of the limits may estimate location at a rate faster than \(n^{-1/2}\) in bounded-support families, as demonstrated by the uniform model. Such gains arise from the relation between the parameter, the support, and the sampling law, not from a universal superiority of extreme observations. The contribution of the window must therefore be derived for the family and target under consideration rather than represented by a fixed heuristic weight.

The complete likelihood already contains both components. Multiplying it by the boundary likelihood again would duplicate information and generally change the estimator. Window refinement instead defines
\[
\ell_{\mathrm{WR}}(\theta;\lambda)
=
\ell_{\mathrm{I}\mid\mathrm{B}}(\theta)
+
\lambda\ell_{\mathrm{B}}(\theta),
\]
with \(\lambda=1\) corresponding to the original experiment. Values \(\lambda\neq1\) deliberately change the estimating criterion. This construction differs from truncating a composite likelihood to remove noisy contributions, as in the procedure of \citet{Huang2021379}: no boundary or internal component is discarded, but their relative inferential contributions are made explicit. It also differs from conditional order-statistic likelihood used to avoid specification of nuisance observation processes~\citep{Shen2019831}, because the boundary likelihood remains part of the proposed criterion.

\subsection{Conditioning, Weighting, and Extrapolation}
\label{subsec:discussion_conditioning_weighting}

Hard-window conditioning changes normalization but preserves every internal ratio:
\[
\frac{
	g_{\mathcal W}(x;\theta)
}{
	g_{\mathcal W}(y;\theta)
}
=
\frac{
	g(x;\theta)
}{
	g(y;\theta)
}.
\]
It follows that conditioning cannot correct an internal residual pattern and cannot distinguish parent models that are proportional throughout the represented domain. This equivalence is an inferential limit rather than a failure of optimization. External support, tail mass, and complete normalization remain unidentified by the normalized internal law.

The role of conditioning depends on what is held fixed and what remains inferentially active. In mixture models, conditioning on ancillary statistics can simplify estimation while sacrificing a quantifiable amount of information~\citep{Yip19912045}. In the present construction, the limits are not removed as ancillary quantities: their joint law defines \(L_{\mathrm{B}}\), while conditioning on them defines \(L_{\mathrm{I}\mid\mathrm{B}}\). The resulting decomposition preserves both sources of information and permits their roles to be examined separately.

Internal weighting has a different effect. If \(v(x)\) varies over the window, then
\[
\frac{
	g_v(x;\theta)
}{
	g_v(y;\theta)
}
=
\frac{
	g(x;\theta)
}{
	g(y;\theta)
}
\frac{
	v(x)
}{
	v(y)
}.
\]
Within the weighted-distribution construction of \citet{PatilRao1978}, the weight may represent detection, retention, accessibility, or utility. Separate interpretation of the baseline and weight nevertheless requires constrained model families because only their normalized product is observed. Boundary reweighting through \(\lambda\) modifies the contribution of the window to estimation; internal weighting through \(v(x)\) modifies relative representation within that window. The two operations are not interchangeable.

Censoring in extreme-value analysis provides a further contrast. Pairwise censored likelihood retains threshold information while estimating space--time extremal dependence~\citep{Huser2014439}, and higher-dimensional censored Poisson-process likelihood incorporates sub-threshold components into inference on spatial extremes~\citep{Wadsworth20141}. These procedures define censoring relative to thresholds chosen for tail inference. The observed window instead arises from the complete sample realization and partitions its information into boundary and conditional internal components.

This distinction also separates estimation from extrapolation. A parameter identified by the conditional configuration describes the represented domain. Probabilities below \(X_{(1)}\) or above \(X_{(n)}\) require transfer of that parameter to a parent law whose continuation has not been observed. Under model misspecification, the complete-support estimator converges to a pseudo-true parameter determined by the complete criterion~\citep{White1982}; that target need not coincide with the parameter that best represents a declared internal domain. Reporting the conditional estimate and the parent-model extrapolation as separate inferential layers makes this dependence explicit.

The contamination experiment illustrates the consequence. The complete Gaussian location estimator accommodated the asymmetric external component and moved away from the core target, whereas conditioning on a central empirical window reduced bias from \(0.6038\) to \(0.1098\) and RMSE from \(0.6472\) to \(0.2543\). This result does not establish uniform dominance of local conditioning. It establishes that complete-support and observed-domain criteria can target different aspects of a distribution when its external continuation is unreliable.

\subsection{Calibration, Multiple Windows, and Scope}
\label{subsec:discussion_calibration_scope}

For \(\lambda\neq1\), the window-refined criterion is not the log-likelihood of the original sample experiment. Its curvature and score variability need not coincide, so uncertainty must be based on
\[
H_\lambda^{-1}
J_\lambda
H_\lambda^{-\mathsf T}
\]
rather than on \(H_\lambda^{-1}\) alone. Composite likelihood has the same inferential requirement because its components do not generally reproduce the information identity~\citep{VarinReidFirth2011}. Adjustments to composite likelihood-ratio statistics are consequently needed to recover an ordinary chi-squared reference distribution~\citep{Pace2011}. In the numerical assessment, inverse-sensitivity intervals covered only \(90.60\%\) of replicates when the boundary contribution was doubled, whereas Godambe-calibrated intervals covered \(94.60\%\).

The same calibration principle applies beyond composite likelihood. Proper-scoring-rule posteriors use the associated Godambe information to account for replacing the full log-likelihood by a criterion selected for robustness or tractability~\citep{Giummole2019}. The window-refined criterion belongs to this broader class when \(\lambda\neq1\): its point estimator is defined by a valid objective, but likelihood-based uncertainty does not follow from curvature alone.

The choice of \(\lambda\) remains part of the inferential specification. Unit weight is fixed by the original sampling experiment. A different value may be derived from an explicit loss, robustness objective, or relative-information criterion. The robustness of a composite estimator depends on the components retained and on their relation to the data-generating structure~\citep{Xu2011}; analogously, boundary reweighting protects a declared target only when the boundary and internal components are aligned with that target. Selection of \(\lambda\) from the same observations introduces additional uncertainty and must be included in the sampling analysis. The present results establish the criterion and its calibration, not a universal value of \(\lambda\).

The multiple-sample formulation similarly distinguishes targets rather than ranking interchangeable estimators. Sample-specific windows preserve the observation domain of each sample. A pooled union treats the combined observations as if they shared the widest domain, while a common intersection restricts comparison to the region represented in every sample. The simulation recovered the common parameter under sample-specific normalization, exposed bias under pooled-union normalization, and showed the variance cost of restriction to a narrow intersection. These differences arise from the observation domains assigned to the samples, not from alternative numerical implementations of the same experiment.

Several extensions follow from the same structure. Discrete and mixed distributions require boundary likelihoods that retain ties and boundary multiplicities. Multivariate samples require a declared region in place of the univariate interval; estimating the shape of a compact limiting point cloud~\citep{Majumder20261191}, for example, illustrates how multivariate boundary geometry may itself become the inferential object. That problem differs from the present use of a realized window, but it confirms that boundary shape can carry parameter information not reducible to marginal extremes.

Raw minima and maxima are sensitive to contamination. Quantile or scientifically specified windows can provide robust alternatives, but they define distinct targets and must not be interpreted as the complete realized domain. Time aggregation introduces a related distinction: observations at several aggregation scales are mathematically constrained because each coarser scale reuses information from finer scales rather than adding independent data~\citep{Ailliot2026}. Likewise, nested or transformed windows must be analyzed through their dependence structure rather than counted as separate samples. Asymptotic development must finally account jointly for parameter estimation and the evolution of the random window, especially under unbounded support.

\section{Conclusions}
\label{sec:conclusions}

The observed sample contains information about both its realized boundaries and its internal configuration. For continuous iid data, these contributions admit the exact factorization
\[
L_{\mathrm{full}}(\theta)
\propto
L_{\mathrm{B}}(\theta)
L_{\mathrm{I}\mid\mathrm{B}}(\theta).
\]
This identity provides a non-duplicative basis for observed-window inference and shows that the limits cannot be treated either as two ordinary observations or as an externally fixed truncation region.

The window-refined criterion
\[
\ell_{\mathrm{WR}}(\theta;\lambda)
=
\ell_{\mathrm{I}\mid\mathrm{B}}(\theta)
+
\lambda\ell_{\mathrm{B}}(\theta)
\]
recovers complete likelihood at \(\lambda=1\) and permits explicit modification of the boundary contribution for other inferential targets. Non-unit weighting requires Godambe calibration. It remains distinct from internal weighting, which changes probability or density ratios within the represented domain.

The framework consequently separates four objects: boundary information, internal configuration, internal weighting, and external continuation. The first two constitute the observed sampling experiment; the third represents outcome-dependent modification inside the window; the fourth supplies model-dependent extrapolation beyond it. Their explicit separation provides a compact basis for distributional estimation when the realized observational domain is itself scientifically or statistically informative.

\section*{Acknowledgements}

The author acknowledges the use of Google Gemini and Microsoft Copilot, powered by GPT-5.6, for assistance with code verification, computational validation, mathematical checking, and language revision. The author determined the methodological formulation, verified the numerical results, interpreted the findings, and accepts responsibility for the manuscript.

\section*{Funding}

No external funding was received for this research.

\section*{Conflict of Interest}

The author declares no conflict of interest.

\section*{Data Availability}

No external empirical dataset was used. The numerical assessment was generated from the probability models and parameter settings stated in the manuscript and supplementary material.

\section*{Supplementary Material}

The supplementary material contains the positioning of the proposed framework within the scientific literature, mathematical specifications of the verification experiments, numerical verification results, and finite-sample assessments of window-refined estimation.

\section*{Author Contributions}

Lorentz J\"antschi conceived the study, developed the methodology, designed and conducted the numerical analysis, interpreted the results, and wrote the manuscript.

\end{document}